\documentclass[journal]{IEEEtran}
\usepackage{cite}
\usepackage{amsmath,amssymb,amsfonts}
\usepackage{graphicx}
\usepackage{textcomp}
\usepackage{caption}
\usepackage{tikz}
\usepackage{mwe} 
\usepackage{subfigure}
\usepackage[subfigure]{tocloft}
\usepackage{float}
\usepackage{subfigure}
\usepackage[ruled,vlined]{algorithm2e}
\usepackage{graphicx}
\usepackage{layout}
\usepackage{wrapfig}
\usepackage{multirow}
\usepackage{booktabs}
\usepackage{amsmath,amssymb}
\usepackage[ruled,vlined]{algorithm2e}
\RestyleAlgo{boxruled}

\begin{document}
\title{Input-Dependent Fisher Information for Local Sensitivity Analysis of Medical Image Classifiers}
\author{Sourya Sengupta and  Mark A. Anastasio \IEEEmembership{Fellow, IEEE}
\thanks{This work was supported in part by NIH Awards EB031772 (subproject 6366), EB031585 and EB034249.  Research reported in this publication was supported by the National Institute Of Biomedical Imaging And Bioengineering of the National Institutes of Health under Award Number T32EB019944. The content is solely the responsibility of the authors and does not necessarily represent the official views of the National Institutes of Health. }
\thanks{Sourya Sengupta is with the Department of Electrical and Computer Engineering, University of Illinois Urbana–Champaign, Urbana,
IL 61801 USA (e-mail: souryas2@illinois.edu).}
\thanks{Mark A. Anastasio is with the Mallinckrodt Institute of Radiology and Department of Electrical \& Systems Engineering, Washington University in St. Louis, St. Louis, MO 63110 USA (e-mail: maa@illinois.edu).}}

\maketitle

\begin{abstract}
Deep neural networks have achieved strong performance in medical image classification, but often work like black-box. Commonly used post-hoc interpretation methods often provide heuristic visualizations whose relationship to the classifier's predictive distribution is indirect. This work introduces a local sensitivity analysis framework based on the input-dependent Fisher Information Matrix (iFIM) of a trained classifier. The iFIM characterizes how the classifier's predictive distribution changes under infinitesimal perturbations of the input image. By using a Gram-matrix formulation, the nonzero eigenspectrum of the iFIM can be recovered without explicitly forming the full image-dimensional Fisher matrix. The leading iFIM eigenspace is then used to project an input image into a high local-sensitivity component and its orthogonal component. These components provide a model-intrinsic description of local predictive sensitivity, rather than a conventional pixel-wise attribution heatmap or a causal segmentation of task-relevant anatomy. The framework is evaluated on controlled and clinical medical image classification tasks using multiple classifier architectures. Perturbation-based experiments show that high-sensitivity iFIM components are more strongly coupled to changes in predictive confidence and classification performance than lower-sensitivity complementary components. The results support the iFIM framework as a principled tool for analyzing local decision sensitivity and for complementing existing attribution-based interpretability methods in medical imaging.
\end{abstract}


\section{Introduction}
Deep learning-based classifiers have achieved strong performance across a wide range of medical imaging tasks, but methods for interpreting their decisions remain limited. Most widely used post-hoc approaches produce attribution or localization maps that indicate which image regions are associated with the model output, but the quantities they visualize are often heuristic and may vary across architectures, implementations, or perturbations. As a result, these methods can be useful for inspection, yet they do not always provide a principled characterization of how the predictive distribution of the classifier responds to changes in the input.


In this work, we propose an information-theoretic framework for analyzing how a classifier’s predictive distribution changes locally under input perturbations.
The framework is based on the input-dependent Fisher Information Matrix (iFIM), which quantifies the local sensitivity of the predictive distribution to input perturbations. By spectrally decomposing the iFIM, the input image is partitioned into components associated with directions of relatively high and low local predictive sensitivity. Henceforth, these are referred to as the task-sensitive and task-insensitive components, respectively, in a relative and local sense. Rather than serving as a replacement for attribution heatmaps, the proposed framework provides a complementary, model-intrinsic representation of how perturbations near a given input influence the classifier output.


The major contributions of this work are as follows:
\begin{itemize}
\item We formulate an input-dependent Fisher Information Matrix framework for characterizing the local sensitivity of a trained medical image classifier's predictive distribution to input perturbations.

\item We use a Gram-matrix factorization to recover the nonzero iFIM eigenspectrum without constructing the full image-dimensional Fisher matrix, enabling tractable computation for image-scale inputs.

\item We define a projection-based decomposition of an input image into a high local-sensitivity component and its orthogonal complement, and clarify the interpretation of this decomposition for low-rank settings such as binary classification.

\item We evaluate the resulting sensitivity components across controlled and clinical medical image classification tasks using perturbation-based analyses and architecture comparisons.
\end{itemize}

\if o
\begin{itemize}
\item A novel information-theoretic framework for analyzing model interpretability is introduced based on the input-dependent Fisher Information Matrix (iFIM). Unlike CAM-based post-hoc methods that rely on intermediate feature activations, iFIM directly quantifies task sensitivity, defined as the local sensitivity of the model’s predictive distribution with respect to the input image. This perspective serves as a complementary geometric analysis rather than a replacement for localization-based interpretability methods.

\item Although iFIM-based analysis is not directly comparable to CAM-based attribution maps, it is evaluated using commonly adopted perturbation-based metrics to contextualize its behavior within the existing interpretability literature. Across architectures, iFIM exhibits stronger responses under task-sensitive perturbation analyses and adversarial perturbations, consistent with its formulation as a measure of local task sensitivity.

\item The iFIM formulation enables an orthogonal decomposition of the input into task-sensitive and task-insensitive components, corresponding to directions of higher and lower local task sensitivity, respectively. This yields interpretable representations of how input variations influence the model’s output in a model-intrinsic manner. This decomposition provides insights into model behavior that complement, rather than replace, conventional localization-based attribution techniques.
\end{itemize}
\fi

\section{Background}

Existing approaches to interpretability in deep learning can be broadly divided into post-hoc attribution methods and inherently interpretable model designs. Post-hoc methods, including gradient saliency maps, SmoothGrad, integrated gradients, and class activation map (CAM) variants such as Grad-CAM, Grad-CAM++, Score-CAM, and Eigen-CAM, are commonly used to visualize image regions associated with a model’s output \cite{tang2024reviewing, Simonyan2014Saliency, Selvaraju2017GradCAM, Chattopadhyay2018GradCAMpp, Wang2020ScoreCAM, Muhammad2020EigenCAM, Smilkov2017SmoothGrad, Bach2015LRP, Shrikumar2017DeepLIFT, Zhou2016CAM, hu2024unioncam, yamauchi2024spatial}. These methods can be useful for qualitative inspection and spatial localization, but the quantities they visualize are often heuristic and may depend on gradient flow, feature-map selection, or implementation choices. Consequently, attribution maps do not always provide a clear or principled characterization of how the predictive distribution of the classifier changes under perturbations of the input.

A different line of work seeks interpretability through architectural design. Examples include concept bottleneck models, self-explaining neural networks, prototype-based networks, and related approaches that constrain the model to expose interpretable intermediate representations \cite{jain2020learning, alvarez2018towards, koh2020concept, chen2019looks, agarwal2021neural, sengupta2024test}. Such methods can provide explanations that are more tightly coupled to model structure, but they typically require architectural modification or additional supervision and may reduce flexibility relative to unconstrained high-capacity classifiers \cite{wang2021self, mohammadjafari2021using}. In many medical imaging settings, it is therefore desirable to have tools that can analyze the behavior of standard trained classifiers without redesigning the predictive model itself.

These considerations motivate the need for a model-intrinsic way to characterize how a trained classifier depends on its input. In this work, we adopt a local view of this dependence by examining how the classifier’s predictive distribution changes under small perturbations of the input. The Fisher Information Matrix (FIM) provides a principled measure of the local sensitivity of a probability distribution to such perturbations. When defined with respect to the input rather than the model parameters, the resulting input-dependent Fisher Information Matrix (iFIM) characterizes the local sensitivity of a classifier’s predictive distribution in input space. This makes the iFIM a natural foundation for analyzing local decision geometry without modifying the classifier architecture. 

Fisher information has been used in deep learning in several other contexts, including natural-gradient-based optimization, spectral and trainability analyses of network parameter spaces, pruning and compression, and robustness analysis \cite{martens2015optimizing, karakida2019universal, hayase2021spectrum,picot2022adversarial}. In contrast, the present work focuses on the iFIM of the predictive distribution and uses its spectral structure to construct an orthogonal decomposition of the input image into components associated with relatively high and low local predictive sensitivity.

In the following section, we show how the spectral decomposition of the iFIM can be used to identify directions of relatively high and low local predictive sensitivity, and thereby to construct a corresponding decomposition of the input image. This perspective complements conventional attribution-based analyses by focusing not on pixel-wise importance scores, but on the local sensitivity structure of the predictive distribution induced by the trained classifier.

\section{Method}
The proposed method takes as input a trained probabilistic classifier and an image, and returns a pair of orthogonal projections of the image associated with different levels of local predictive sensitivity. Specifically, the input-dependent Fisher Information Matrix (iFIM) of the classifier's predictive distribution is evaluated at the image, its leading nonzero eigenvectors are computed through a tractable Gram-matrix formulation, and the image is projected onto the corresponding high-sensitivity eigenspace and its orthogonal component. These projected components provide a local sensitivity-based representation of model behavior near the given image. They should be interpreted as projections onto directions of higher and lower local sensitivity of the predictive distribution, rather than as a spatial partition of semantically relevant and irrelevant anatomy.

\subsection{Input-Dependent Fisher Information and Gram Factorization}

Let \(x\in\mathbb{R}^d\) denote an input image represented as a vector, and let
\(y\in\{1,\ldots,C\}\) denote the class label. Consider a trained classifier
\(f_\theta:\mathbb{R}^d\rightarrow\mathbb{R}^C\), where \(f_\theta(x)_c\) is the
logit for class \(c\). The predictive distribution is
\begin{equation}
p_\theta(y=c\mid x)
=
\frac{\exp(f_\theta(x)_c)}
{\sum_{j=1}^{C}\exp(f_\theta(x)_j)} .
\end{equation}

To characterize local changes in this predictive distribution with respect to the input, we define
the input-dependent Fisher Information Matrix (iFIM) as
\begin{equation}
\mathbf{F}(x)
=
\mathbb{E}_{y\sim p_\theta(\cdot\mid x)}
\left[
\nabla_x \log p_\theta(y\mid x)
\nabla_x \log p_\theta(y\mid x)^\top
\right].
\end{equation}
The iFIM is symmetric positive semidefinite and characterizes the local sensitivity of the
predictive distribution to infinitesimal input perturbations.

Let
\begin{equation}
g_c(x)=\nabla_x \log p_\theta(y=c\mid x).
\end{equation}
Then the iFIM can be written as
\begin{equation}
\mathbf{F}(x)
=
\sum_{c=1}^{C}
p_\theta(y=c\mid x)\, g_c(x)g_c(x)^\top .
\end{equation}
Define
\begin{equation}
\begin{aligned}
\mathbf{G}(x)
&=
\Big[
\sqrt{p_\theta(y=1\mid x)}\,g_1(x),\dots, \\
&\qquad
\sqrt{p_\theta(y=C\mid x)}\,g_C(x)
\Big]
\in\mathbb{R}^{d\times C}.
\end{aligned}
\label{eq:G_matrix}
\end{equation}
With this definition,
\begin{equation}
\mathbf{F}(x)=\mathbf{G}(x)\mathbf{G}(x)^\top .
\end{equation}

Directly constructing \(\mathbf{F}(x)\in\mathbb{R}^{d\times d}\) is generally impractical for
image-scale inputs. We therefore work with the smaller Gram matrix
\begin{equation}
\mathbf{K}(x)=\mathbf{G}(x)^\top\mathbf{G}(x)\in\mathbb{R}^{C\times C}.
\end{equation}
Because \(\mathbf{F}(x)=\mathbf{G}(x)\mathbf{G}(x)^\top\) and
\(\mathbf{K}(x)=\mathbf{G}(x)^\top\mathbf{G}(x)\), their nonzero eigenvalues coincide. Let
\begin{equation}
\mathbf{K}(x)\mathbf{v}_i=\lambda_i\mathbf{v}_i,
\qquad
\lambda_i>0,
\end{equation}
where \(\lambda_1\ge\lambda_2\ge\cdots\ge\lambda_R>0\), the eigenvectors
\(\{\mathbf{v}_i\}_{i=1}^R\) are orthonormal, and
\(R=\mathrm{rank}(\mathbf{F}(x))\le C-1\). The corresponding unit-norm eigenvector of
\(\mathbf{F}(x)\) is
\begin{equation}
\mathbf{u}_i
=
\frac{\mathbf{G}(x)\mathbf{v}_i}{\sqrt{\lambda_i}} .
\end{equation}
Collecting the nonzero eigenvectors as
\(\mathbf{U}_R=[\mathbf{u}_1,\ldots,\mathbf{u}_R]\), with
\(\mathbf{V}=[\mathbf{v}_1,\ldots,\mathbf{v}_R]\) and
\(\boldsymbol{\Lambda}=\mathrm{diag}(\lambda_1,\ldots,\lambda_R)\), gives
\begin{equation}
\mathbf{U}_R
=
\mathbf{G}(x)\mathbf{V}\boldsymbol{\Lambda}^{-1/2},
\end{equation}
and hence the nonzero eigendecomposition
\begin{equation}
\mathbf{F}(x)
=
\mathbf{U}_R\boldsymbol{\Lambda}\mathbf{U}_R^\top .
\end{equation}

As an illustrative example, consider the binary classification case. When \(C=2\), define the
logit-difference function
\[
s(x)=f_\theta(x)_1-f_\theta(x)_2,
\]
and let \(p_1=p_\theta(y=1\mid x)\) and \(p_2=p_\theta(y=2\mid x)\). Then
\[
\nabla_x \log p_1 = p_2\nabla_x s(x),
\qquad
\nabla_x \log p_2 = -p_1\nabla_x s(x),
\]
and the iFIM becomes
\begin{equation}
\mathbf{F}(x)
=
p_1p_2\,
\nabla_x s(x)\nabla_x s(x)^\top .
\end{equation}
Thus, in the binary softmax case, the iFIM has rank one, and its nonzero eigendirection
is aligned with the local sensitivity direction of the logit difference.

The iFIM and the associated Gram-matrix factorization provide the basis for the image decomposition
introduced next.

\subsection{iFIM-Based Image Decomposition}

The methodological contribution of this work is to use the spectral structure of the iFIM to define
an orthogonal projection of the input image onto subspaces associated with different levels of local
predictive sensitivity. 

Let \(\{u_i\}_{i=1}^{d}\) denote an orthonormal eigenbasis of \(\mathbf{F}(x)\), with eigenvalues
ordered as
\[
\lambda_1\ge \lambda_2\ge \cdots \ge \lambda_R>0,
\qquad
\lambda_{R+1}=\cdots=\lambda_d=0 .
\]
The eigenvectors associated with the nonzero eigenvalues are given by the columns of
\(\mathbf{U}_R\), while the remaining eigenvectors span the nullspace of \(\mathbf{F}(x)\).

We select the smallest \(K\le R\) such that
\begin{equation}
\frac{\sum_{i=1}^{K}\lambda_i}{\sum_{i=1}^{R}\lambda_i}\ge \rho,
\end{equation}
where \(\rho\in(0,1]\) is a prescribed energy threshold. Let
\[
\mathbf{U}_K=[u_1,\ldots,u_K].
\]
The high-sensitivity projection is defined as
\begin{equation}
\mathbf{P}_{\mathrm{sens}}
=
\mathbf{U}_K\mathbf{U}_K^\top
=
\sum_{i=1}^{K}u_i u_i^\top ,
\end{equation}
while the complementary projection is
\begin{equation}
\mathbf{P}_{\mathrm{comp}}
=
\mathbf{I}-\mathbf{P}_{\mathrm{sens}} .
\end{equation}
Accordingly, the input image can be decomposed as $x = x_{\mathrm{sens}} + x_{\mathrm{comp}}$, where
\begin{equation}
\qquad
x_{\mathrm{sens}}=\mathbf{P}_{\mathrm{sens}}x,
\quad
x_{\mathrm{comp}}=\mathbf{P}_{\mathrm{comp}}x.
\end{equation}
By construction, \(x_{\mathrm{sens}}\) and \(x_{\mathrm{comp}}\) are orthogonal.

The component \(x_{\mathrm{sens}}\) lies in the span of the leading iFIM eigenvectors and is therefore
associated with directions of relatively high local predictive sensitivity. The complementary
component \(x_{\mathrm{comp}}\) contains both lower-sensitivity nonzero eigendirections and, when
present, directions in the nullspace of \(\mathbf{F}(x)\). Thus, \(x_{\mathrm{comp}}\) should be
interpreted as the component orthogonal to the selected high-sensitivity subspace, rather than as a
collection of globally irrelevant or causally uninvolved image features.

Although the decomposition is defined in the full input space, computing \(x_{\mathrm{sens}}\) only
requires the leading \(K\) eigenvectors recovered from the Gram-matrix formulation. The complementary
component is then obtained as
\begin{equation}
x_{\mathrm{comp}}=x-x_{\mathrm{sens}},
\end{equation}
without explicitly computing the remaining \(d-K\) eigenvectors. For visualization and
pixel-perturbation experiments, the projected components are reshaped to the spatial dimensions of
the input image.

\subsection{Local Perturbation Interpretation of the iFIM}

Let \(p_\theta(\cdot\mid x)\) denote the predictive distribution of a trained classifier at input
\(x\), and let \(\mathbf{F}(x)\) be the corresponding iFIM. For a small perturbation
\(\delta\in\mathbb{R}^d\), local changes in the predictive distribution can be quantified by
\[
\mathrm{KL}\!\left(p_\theta(\cdot\mid x)\,\|\,p_\theta(\cdot\mid x+\delta)\right).
\]
At points where the predictive distribution is locally twice differentiable with respect to the
input, a second-order Taylor expansion gives
\begin{equation}
\mathrm{KL}\!\left(p_\theta(\cdot\mid x)\,\|\,p_\theta(\cdot\mid x+\delta)\right)
=
\frac{1}{2}\,\delta^\top \mathbf{F}(x)\,\delta
+
o(\|\delta\|_2^2).
\label{eq:kl_quadratic}
\end{equation}
Thus, to leading order, the iFIM determines the local sensitivity of the predictive distribution to
input perturbations \cite{amari2000methods}.

Let
\begin{equation}
\mathbf{F}(x)=\mathbf{U}\boldsymbol{\Lambda}\mathbf{U}^\top
\end{equation}
be a full eigendecomposition of the symmetric positive semidefinite iFIM, where
\(\mathbf{U}=[u_1,\ldots,u_d]\) is an orthonormal basis and
\(\boldsymbol{\Lambda}=\mathrm{diag}(\lambda_1,\ldots,\lambda_d)\), with
\(\lambda_1\ge\lambda_2\ge\cdots\ge\lambda_d\ge0\). Although this full basis is useful for
interpretation, only the nonzero eigenvectors need to be computed in practice using the
Gram-matrix formulation described above.

Any perturbation can be written as
\begin{equation}
\delta=\sum_{i=1}^{d} a_i u_i,
\qquad
a_i=u_i^\top\delta,
\qquad
\|\delta\|_2^2=\sum_{i=1}^{d} a_i^2.
\end{equation}
Since \(\mathbf{U}^\top\delta=(a_1,\ldots,a_d)^\top\), the quadratic form becomes
\begin{equation}
\delta^\top \mathbf{F}(x)\delta
=
\sum_{i=1}^{d}\lambda_i a_i^2.
\label{eq:ifim_spectral}
\end{equation}
Therefore, the local KL sensitivity decomposes into orthogonal contributions weighted by the iFIM
eigenvalues. Directions in the nullspace of \(\mathbf{F}(x)\), for which \(\lambda_i=0\), do not
contribute to this second-order approximation.

If \(\|\delta\|_2\le\epsilon\), then
\begin{equation}
\delta^\top \mathbf{F}(x)\delta
\le
\lambda_1 \sum_{i=1}^{d} a_i^2
\le
\lambda_1\epsilon^2.
\end{equation}
Equality is achieved by \(\delta=\epsilon u_1\) when the leading eigenvalue is simple, or more
generally by any perturbation of norm \(\epsilon\) lying in the eigenspace associated with
\(\lambda_1\). 
Thus, under an \(\ell_2\)-norm constraint, the leading iFIM eigenspace gives the direction or directions that maximize the leading-order change in the predictive distribution. Equivalently, among perturbations of equal norm, larger iFIM eigenvalues correspond to greater local predictive sensitivity.

This interpretation is strictly local. It orders perturbation directions according to their
leading-order effect on the predictive distribution near \(x\), but it does not imply a global,
semantic, or causal ordering of image features. For finite perturbations, directions with larger
projections onto high-eigenvalue eigenspaces are expected to produce larger changes in the predictive
distribution when perturbations are sufficiently small. However, changes in confidence or
classification performance need not be monotonic, because they also depend on perturbation sign,
class margins, and nonlinear classifier behavior beyond the local quadratic approximation.

\subsection{Computational Complexity of iFIM-map Computation}

Direct construction of the full iFIM \(\mathbf{F}(x)\in\mathbb{R}^{d\times d}\) is infeasible for
image-scale inputs because it requires storing \(O(d^2)\) entries and performing spectral
computations in the ambient input dimension. The proposed formulation avoids this by working with
\(\mathbf{G}(x)\in\mathbb{R}^{d\times C}\) and the corresponding Gram matrix
\(\mathbf{K}(x)=\mathbf{G}(x)^\top\mathbf{G}(x)\in\mathbb{R}^{C\times C}\).

The dominant cost is the computation of the class-wise gradients
\(g_c(x)=\nabla_x \log p_\theta(c\mid x)\), which requires one input-gradient evaluation per class.
Once these gradients are available, forming \(\mathbf{K}(x)\) requires \(O(dC^2)\) operations and
its eigendecomposition requires \(O(C^3)\), which is negligible when \(C\) is small. In typical
medical imaging classification problems, where \(d\) may be on the order of \(10^4\)–\(10^6\) and
\(C\) is often relatively small ($C=2$ for binary tasks), the Gram-matrix formulation makes the spectral step tractable while
preserving the nonzero spectral information of the iFIM. 

\subsection{Relation to Post-hoc Attribution Methods}

The proposed framework differs from conventional post-hoc attribution methods in the quantity it
seeks to characterize. Attribution methods typically produce pixel-wise importance or localization
maps derived from gradients, feature activations, or related heuristics. These maps are primarily
used to visualize spatial regions that influence a model's output.

In contrast, the iFIM-based analysis characterizes the local sensitivity structure of the predictive
distribution in input space. Its output is therefore not a conventional attribution heatmap, but a
set of orthogonal directions and projected image components associated with relatively high and low
local predictive sensitivity. The task-sensitive and task-insensitive components defined in this
work should accordingly be interpreted as components of local predictive sensitivity, rather than as
definitive segmentations of semantically or causally relevant image content.

For this reason, the proposed framework is best viewed as complementary to localization-based
explanation methods rather than as a direct replacement for them. It provides a geometric
description of how perturbations near a given input affect the predictive distribution of the trained
classifier, whereas standard attribution methods are primarily designed to visualize spatial regions
associated with model output. A comparison between these perspectives and conventional post-hoc
attribution methods is summarized in Table~\ref{tab:ifim_vs_posthoc}.

\begin{table}[t]
\centering
\caption{Comparison between conventional post-hoc attribution methods and the proposed iFIM-based analysis framework.}
\label{tab:ifim_vs_posthoc}
\resizebox{\columnwidth}{!}{
\begin{tabular}{|p{3.35cm}|p{3.65cm}|}
\hline
\textbf{Post-hoc Attribution Methods} & \textbf{iFIM-based Analysis} \\ \hline

Typically output pixel-wise importance or localization maps &
Outputs orthogonal directions and projected image components associated with relatively high and low local predictive sensitivity \\ \hline

Often derived from gradients, feature activations, or related heuristics &
Derived from the input-dependent Fisher Information Matrix of the predictive distribution \\ \hline

Primarily used for spatial localization and qualitative visualization &
Primarily used to analyze the local sensitivity structure of the predictive distribution in input space \\ \hline

Interpretation varies with the design of the attribution method &
Interpretation is theoretically grounded and describes the local perturbation sensitivity of the predictive distribution \\ \hline

Provides a spatial view of regions associated with model output &
Provides a geometric view of how perturbations near a given input affect model output \\ \hline
\end{tabular}
}
\end{table}

\section{Numerical Studies}

This section evaluates the proposed framework as a method for analyzing local predictive sensitivity in image classifiers. The experiments are designed to demonstrate that the iFIM-based decomposition identifies image components that are more strongly coupled to perturbation-induced changes in the predictive distribution and classifier performance. To this end, we consider controlled and real-world medical imaging tasks, examine qualitative decompositions across architectures, and evaluate the resulting task-sensitive and task-insensitive components using perturbation-based protocols. These studies are intended to assess local sensitivity and model behavior, rather than to establish a complete semantic or causal account of task-relevant image content.

\subsection{Tasks and Datasets}
\label{subsec:tasks}

Three distinct classification tasks were examined in this study.

\textbf{Normal vs. drusen vs. CNV classification using retinal OCT images:}
A three-class classification problem involving normal retina, Drusen, and choroidal neovascularization (CNV) was conducted using optical coherence tomography (OCT) images of the human retina with spatial dimensions of $256 \times 256$ pixels \cite{kermany2018identifying}. Drusen corresponds to the accumulation of extracellular material between the retinal pigment epithelium (RPE) layer and Bruch's membrane in the human retina. These deposits can be clearly visualized in retinal OCT scans. The training dataset consisted of 2000 images from each class, the validation and test sets each contained 100 images per class.

\textbf{Cardiomegaly detection using chest X-ray images:}
A cardiomegaly detection task was performed using chest X-ray images of size $1024 \times 1024$. Cardiomegaly refers to an enlargement of the heart and is commonly used as an indicator of underlying cardiovascular disease. The images were obtained from a publicly available NIH dataset \cite{wang2017chestx}. Image labels were generated through text-mining techniques applied to radiology reports written by clinicians. The training data consisted of 800 images from each class. The validation data consisted of 50 images from each class, and the test data consisted of 70 images from each class.

\textbf{SKS/BKS task with MRI:}
The images were generated using a stylized MRI simulation under signal-known-statistically/background-known-statistically (SKS/BKS) conditions. Background images were derived from the Human Connectome Project dataset \cite{hcp_young_adult_1200}. Gaussian signals with amplitudes ranging from 0.1 to 0.35 were randomly embedded within white matter regions to create signal-present images, with each image containing at most one signal. The resulting dataset consisted of 11,130 images in total, including 8,904 training samples, 1,113 validation samples, and 1,113 testing samples. Each image had spatial dimensions of $256 \times 256$ pixels.


\subsection{Classifier Performance}

For all classification tasks, two backbone classifiers were employed: VGG16 and ResNet18. Each model was trained under the same protocol with cross-entropy loss, the Adam optimizer \cite{kingma2014adam}, and early stopping based on validation performance. Their evaluation metrics are summarized in Table~\ref{tab:precision_recall_three_tasks}. Having established that the classifiers achieve reasonable task performance, the following analyses examine whether the iFIM decomposition reveals image components that are more strongly coupled to perturbation-induced changes in model behavior.

Models were trained with an initial learning rate of \(3\times10^{-5}\), and a learning-rate scheduler (step size \(=8\), decay factor \(=0.9\)). The model checkpoint with the lowest validation loss after epoch 10 was retained. Each experiment was repeated over 10 random initializations. Unless otherwise stated, reported values are mean \(\pm\) standard deviation across these runs.

\begin{table*}[] 
\centering \normalsize \setlength{\tabcolsep}{3pt} \renewcommand{\arraystretch}{0.9}
\caption{Mean $\pm$ standard deviation of precision and recall across three tasks.}
\label{tab:precision_recall_three_tasks} 
\begin{tabular}{lcc|cc|cc} \toprule 
& \multicolumn{2}{c|}{X-ray} 
& \multicolumn{2}{c|}{OCT} 
& \multicolumn{2}{c}{MRI} \\ 
Model 
& Precision & Recall 
& Precision & Recall 
& Precision & Recall \\ 
\midrule 
VGG16 
& $0.77\pm0.03$ & $0.78\pm0.03$ 
& $0.98\pm0.01$ & $0.98\pm0.01$ 
& $0.97\pm0.01$ & $0.96\pm0.01$ \\ 
ResNet18 
& $0.80\pm0.01$ & $0.79\pm0.03$ 
& $0.98\pm0.01$ & $0.98\pm0.01$ 
& $0.97\pm0.01$ & $0.97\pm0.01$ \\ 
\bottomrule 
\end{tabular} 
\end{table*}

\subsection{Visualization of iFIM-based sensitivity components}
\begin{figure}[]
  \centering
  \includegraphics[width=\linewidth]{ 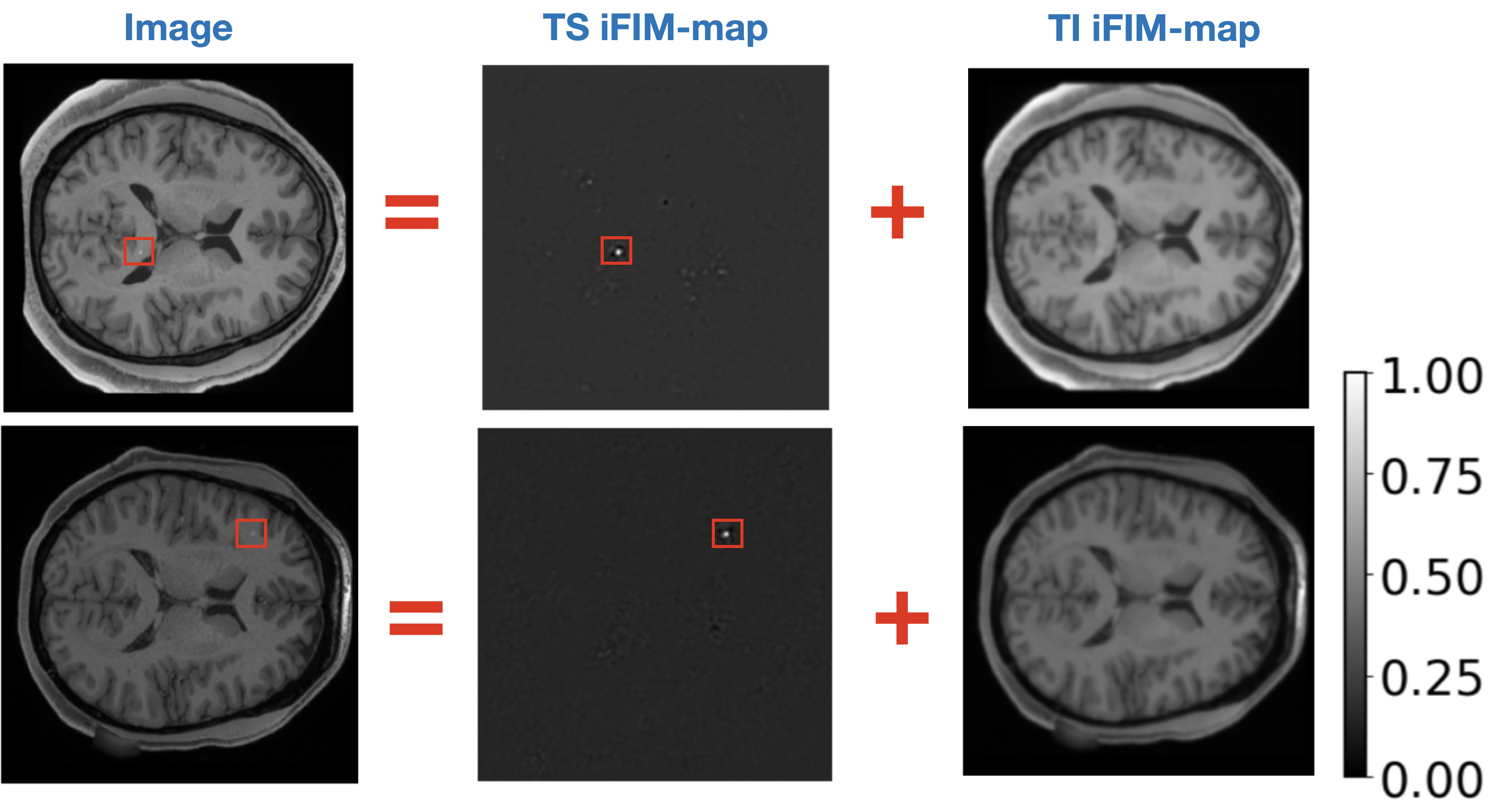}
  \caption{Task-sensitive and task-insensitive iFIM decomposition for a shallow CNN in a controlled MRI setting with a spatially localized signal. The task-sensitive component concentrates around the true signal region, while the task-insensitive component captures background variations. Red bounding boxes show the signal location.}
  \label{fig:cnn_maps}
\end{figure}

Before presenting results on the more complex clinical tasks, we first considered a stylized SKS/BKS MRI task using a shallow convolutional neural network composed of three convolutional layers with ReLU activations. The task-sensitive and task-insensitive iFIM-maps for examples of input images are shown in Fig.~\ref{fig:cnn_maps}. In this controlled setting, the task-sensitive component is concentrated near the known signal region, while the task-insensitive component predominantly captures background variation. This provides an intuitive baseline example in which the iFIM decomposition aligns with a known signal structure under conditions where the target is spatially localized, and the classifier is relatively simple. Task-sensitive iFIM-maps of the CNN for the remaining two tasks are shown in Appendix \ref{sec:AppendixA}.


Figure \ref{fig:all_maps} presents representative task-sensitive and task-insensitive iFIM-maps for deeper architectures, namely VGG16 and ResNet18, on the OCT and chest X-ray tasks. In representative retinal OCT examples (Drusen, CNV), the task-sensitive maps visually align with pathology-bearing retinal layers, including drusen-associated elevations at the retinal pigment epithelium/Bruch’s membrane complex and neovascular sub-/intra-retinal alterations. In chest X-ray cardiomegaly, the task-sensitive maps highlight regions near the enlarged cardiac silhouette. The task-insensitive component can retain pixels within pathology-bearing regions, consistent with the fact that not all pixels in such regions necessarily lie in directions of highest local predictive sensitivity. For both VGG16 and ResNet18, the mean iFIM-map from 10 different runs of the model is shown.

\begin{figure*}[]
  \centering
  \includegraphics[width=\linewidth]{ 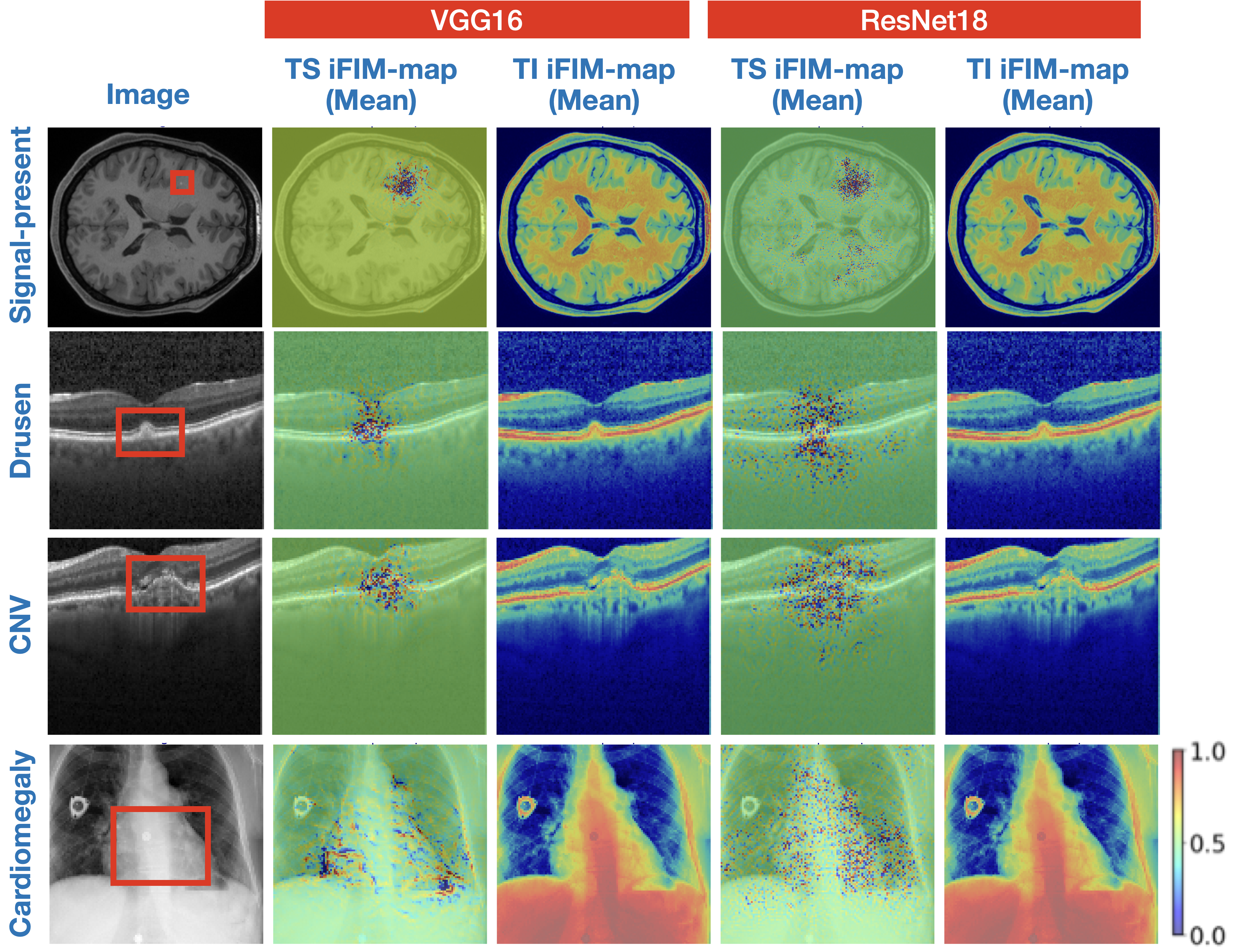}
  \caption{Task-sensitive and task-insensitive iFIM-maps for deeper architectures (VGG16 and ResNet18) across MRI, OCT and chest X-ray tasks. The results are averaged over 10 independent runs. The maps are overlaid on the original images using a contrast-adjusted jet colormap for visualization.}
  \label{fig:all_maps}
\end{figure*}

\subsection{Dependence  on classifier architecture}

A key aspect of the proposed framework is that the resulting decomposition can depend on the classifier used to model the predictive distribution. This dependence arises because the iFIM captures the local sensitivity structure induced by the classifier, which in turn is shaped by the architecture and training dynamics of the model. Fig.~\ref{fig:all_maps} illustrates representative task-sensitive and task-insensitive iFIM-maps obtained using two different classifiers, VGG16 and ResNet18, for cardiomegaly detection from chest X-ray images.

Although both models achieve comparable predictive performance, the spatial distribution of the high-sensitivity component differs across architectures. In the representative examples, the VGG16-based component is more concentrated near the cardiac borders, whereas the ResNet18-based component is more spatially distributed. These observations indicate that iFIM-based projections can reveal architecture-dependent differences in local predictive sensitivity, even when competing classifiers achieve similar aggregate performance. However, these differences should be interpreted as differences in local model behavior rather than as direct evidence that one architecture uses more clinically appropriate features.


\subsection{Impact of Contrastive Pixel Perturbations}

\begin{figure}[]
  \centering
  \includegraphics[width=\linewidth]{ 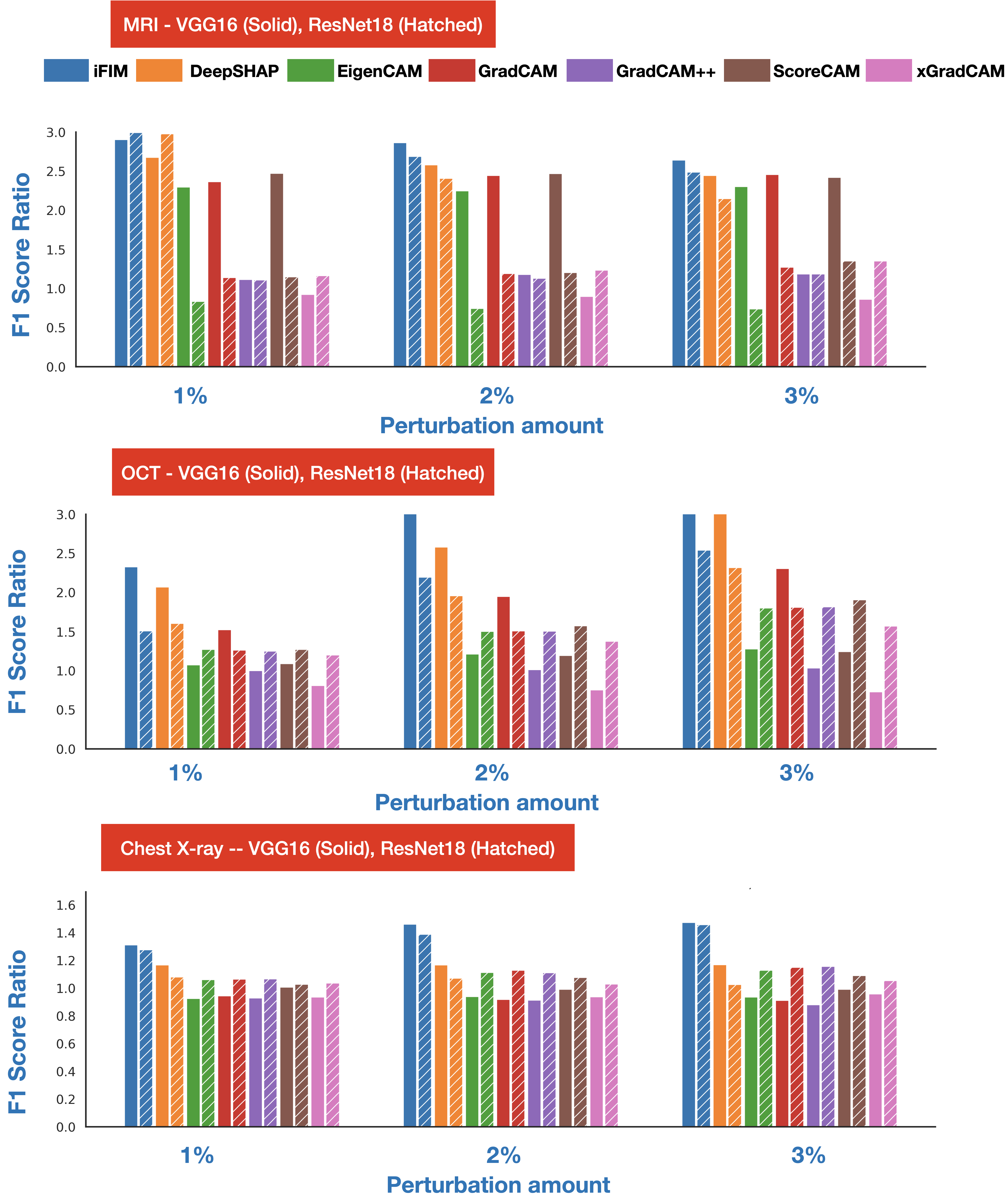}
  \caption{Contrastive pixel-perturbation analysis comparing perturbations selected from the task-sensitive iFIM-map with perturbations selected from the complementary task-insensitive iFIM-map. The plotted value is the ratio of F1 score after task-insensitive iFIM-map perturbation to F1 score after task-sensitive iFIM-map perturbation. Larger values indicate greater performance degradation when perturbing pixels selected from the high-sensitivity component.}
  \label{fig:lerf-morf}
\end{figure}

Sensitivity under targeted pixel perturbation was evaluated using a contrastive perturbation protocol based on the task-sensitive iFIM-map and task-insensitive iFIM-map. Let \(S(x)\in\mathbb{R}^{H\times W}\) denote the task-sensitive iFIM-map and \(I(x)\in\mathbb{R}^{H\times W}\) denote the task-insensitive iFIM-map for input image \(x\). For a perturbation fraction \(\alpha\in(0,1]\), define
\[
k
\equiv \left\lceil \alpha HW \right\rceil .
\]
The task-sensitive iFIM-map perturbation set was selected as the \(k\) pixels with the larger magnitude in \(S(x)\), while the task-insensitive iFIM-map perturbation set was selected as the \(k\) pixels with smaller magnitude in \(I(x)\):
\begin{align}
\mathcal{K}^{S}_{\mathrm{high}}(x,\alpha)
&= \operatorname{TopK}\!\left(|S(x)|,k\right),\\
\mathcal{K}^{I}_{\mathrm{low}}(x,\alpha)
&= \operatorname{BottomK}\!\left(|I(x)|,k\right).
\end{align}
The corresponding perturbed images were constructed as
\begin{align}
\tilde{x}^{S}_{\mathrm{high}}(p)
&=
\begin{cases}
\eta(p), & p\in \mathcal{K}^{S}_{\mathrm{high}}(x,\alpha),\\
x(p), & \text{otherwise},
\end{cases}\\
\tilde{x}^{I}_{\mathrm{low}}(p)
&=
\begin{cases}
\eta(p), & p\in \mathcal{K}^{I}_{\mathrm{low}}(x,\alpha),\\
x(p), & \text{otherwise}.
\end{cases}
\end{align}

where $p$ indexes spatial pixel locations and $\eta(p)$ denotes the perturbation value (e.g., a random draw from a noise distribution). The same pixel index sets are used consistently for perturbation at the specified fraction $\alpha$. The two sets of perturbed images were passed through the trained model, and the resulting F1 scores were compared with the F1 score obtained on the unperturbed images. The F1 scores obtained after perturbing the task-insensitive and task-sensitive pixel sets were computed separately, and their ratio, defined as the task-insensitive perturbation F1 score divided by the task-sensitive perturbation F1 score, was used as the evaluation metric.

A method that more accurately captures the sensitivity of the predictive distribution is expected to yield a higher F1-score ratio, because perturbing its highest-ranked pixels should cause a large reduction in F1 score, whereas perturbing its task-insensitive pixels should produce a substantially smaller reduction.
In this work, $1\%$, $2\%$, and $3\%$ of image pixels were selected according to each attribution map, and the corresponding pixel values were replaced with Gaussian noise drawn from a fixed distribution. For each perturbation level, the performance degradation caused by perturbing high-magnitude entries of the task-sensitive iFIM-map was compared with the degradation caused by perturbing low-magnitude entries of the complementary task-insensitive iFIM-maps. This contrastive protocol evaluates whether the two projected components identify pixel sets with different effects on classifier performance. A comparative analysis of the same has also been performed with other post-hoc attribution maps of DeepSHAP, EigenCAM, GradCAM, GradCAM++, ScoreCAM, xGradCAM as shown in Fig.\ \ref{fig:lerf-morf}. For these methods without an explicit sensitive–insensitive decomposition, the highest- and lowest-ranked pixel locations were selected from the same attribution map.
As shown in Fig.~\ref{fig:lerf-morf}, the iFIM achieved the highest ratio among all evaluated methods across all classifiers and tasks.

\subsection{Impact of adversarial perturbations}

\begin{figure}[]
  \centering
  \includegraphics[width=\linewidth]{ 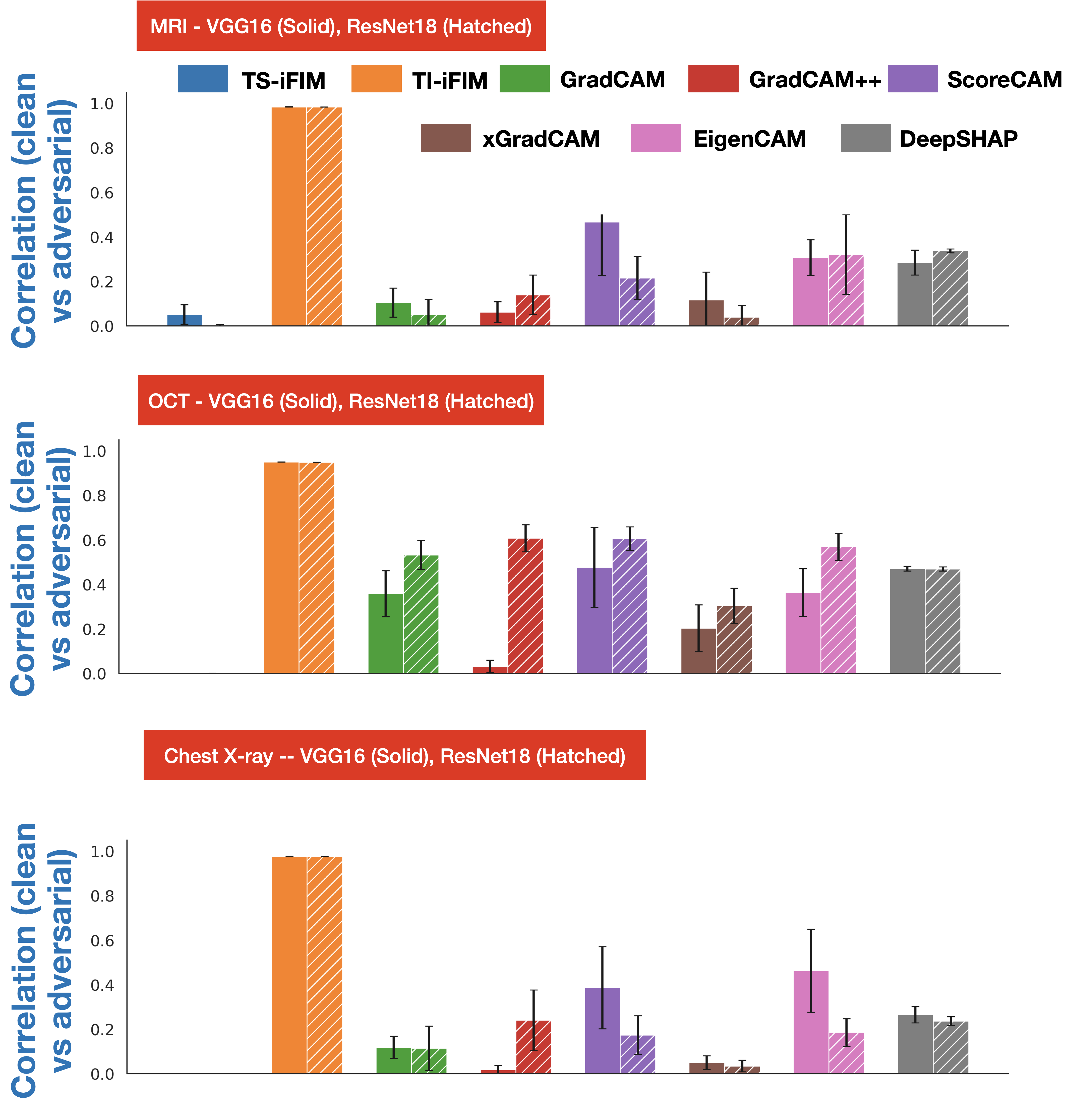}
  \caption{Correlation between maps computed from clean and adversarially perturbed inputs. Lower correlation indicates greater change in the representation under the specified adversarial perturbation. Task-sensitive iFIM-map showed the lowest correlation among all, especially for chest X-ray and OCT tasks, the correlation was almost close to 0 for task-sensitive iFIM-map.
  }
  \label{fig:adversarial}
\end{figure}

To evaluate how strongly different representations respond to decision-altering perturbations, the sensitivity of different representations to adversarial perturbations was examined. Prior work has shown that explanation and saliency representations can exhibit substantial sensitivity to adversarial perturbations, motivating the study of adversarial robustness and stability in interpretability methods \cite{ghorbani2019interpretation,etmann2019connection}. Let $f_{\theta}(x)$ denote a trained classifier with parameters $\theta$, input image $x \in \mathbb{R}^d$, and loss function $\mathcal{L}(f_{\theta}(x), y)$ with respect to ground-truth label $y$. For each test sample, an adversarial example was generated using the Fast Gradient Sign Method (FGSM) \cite{goodfellow2014explaining}, defined as
\begin{equation}
x^{\text{adv}} = x + \epsilon \, \mathrm{sign}\!\left( \nabla_{x} \mathcal{L}(f_{\theta}(x), y) \right),
\end{equation}
where $\epsilon$ controls the perturbation magnitude and $\mathrm{sign}(\cdot)$ is applied elementwise. This formulation corresponds to a first-order approximation of the loss, maximizing $\mathcal{L}$ within an $\ell_{\infty}$-ball of radius $\epsilon$ around $x$.

For each interpretability method, maps were computed for both the clean and adversarially perturbed inputs, denoted by \(M(x)\) and \(M(x^{\mathrm{adv}})\), respectively. The Pearson correlation coefficient between the clean and adversarial maps was then computed.

Mean correlations over the test set were reported for the task-sensitive iFIM-map, the complementary task-insensitive iFIM-map, and competing post-hoc methods. Lower correlation indicates that the corresponding representation changes more strongly under the specified adversarial perturbation. This experiment assesses how strongly each representation changes under perturbations that alter the classifier input in an adversarial direction. Because the iFIM is defined from the local predictive distribution and characterizes local predictive sensitivity, perturbations that alter the classifier’s local decision geometry are expected to affect the iFIM-derived representations. However, low clean/adversarial correlation should be interpreted as sensitivity to the specified perturbation, not as standalone evidence of semantic or causal faithfulness. As shown in Fig.~\ref{fig:adversarial}, the task-sensitive iFIM-map exhibited the lowest clean/adversarial correlation than the compared methods across the examined tasks and architectures. For OCT and Chest X-ray task correlation was almost close to 0 for the task-sensitive iFIM.


Taken together, the results support the proposed framework as a principled method for analyzing local predictive sensitivity in image classifiers. In the controlled MRI testbed, the decomposition aligns with the known signal structure, while in the clinical imaging tasks it yields task-sensitive components that are more strongly coupled to perturbation-induced changes in predictive confidence and classification performance than the compared attribution baselines. The architecture-dependent differences in the resulting decompositions further indicate that the framework can reveal aspects of model behavior that are not captured by aggregate performance metrics alone. At the same time, these experiments primarily support a local sensitivity-based interpretation of the method; stronger claims regarding semantic or causal task relevance would require additional validation against independent reference standards, such as expert annotations or controlled ground-truth localization analyses.
A qualitative comparison with post-hoc attribution methods has also been shown in Appendix B.

\section{Discussion and Conclusion}

This work presents an iFIM-based framework for analyzing the local predictive sensitivity of medical image classifiers. By exploiting the spectral structure of the input-dependent Fisher information matrix, the proposed method projects an input image onto a high-sensitivity eigenspace and its orthogonal component. The resulting components characterize how the predictive distribution of a trained classifier responds locally to perturbations around a given input image. Thus, the framework is best viewed as a local Fisher-geometric analysis tool that complements conventional post-hoc attribution methods.

A key feature of the proposed framework is that it is tied to a single model-intrinsic quantity: the iFIM of the predictive distribution. Unlike many post-hoc attribution methods, which may require method-specific choices such as layer selection, activation weighting, or attribution variants, the iFIM-based analysis follows a uniform computational procedure across classifier architectures. This makes it useful for comparing local sensitivity structure across trained models, even when those models achieve similar aggregate performance.

The numerical studies support the utility of the framework as a local sensitivity analysis method. In the controlled MRI experiment, the high-sensitivity component was concentrated near the known signal region, while the complementary component primarily captured background variation. In the OCT and chest X-ray tasks, the high-sensitivity components were more strongly associated with perturbation-induced changes in classifier performance than the complementary lower-sensitivity components and the compared attribution baselines. The results also showed that iFIM-derived components can differ across classifier architectures, suggesting that the framework can reveal differences in local model behavior that are not captured by aggregate performance metrics alone.

The relevance of local perturbation sensitivity for per-sample interpretation is that a trained classifier defines a local response surface around every input image. Although the observed image itself is fixed, hypothetical perturbations probe how strongly the current predictive distribution depends on different directions in input space. In this sense, the proposed framework does not model actual physical variation in the image, but rather analyzes the local behavior of the classifier at that image. The projected components therefore provide a principled local notion of model dependence, while remaining distinct from stronger claims about global, semantic, or causal task relevance.

Beyond the experiments considered here, the framework may enable several useful directions for studying medical image classifiers. It can be used to investigate whether different architectures rely on different locally sensitive image components, whether robustness interventions alter the local sensitivity structure of the predictive distribution, and whether high-sensitivity components align with pathology-bearing regions or instead reflect spurious cues. Future work should pursue such validation and extend the framework to additional medical imaging tasks, more recent architectures including transformer-based and hybrid models, and multimodal classifiers, where iFIM-based analysis may help characterize how local predictive sensitivity is distributed across input modalities.

\section*{Acknowledgments}
This work was supported in part by NIH Awards P41EB031772 (sub-project 6366), R01EB034249, R01CA233873, R01CA287778, and R56DE033344. The authors would like to acknowledge useful discussions with Drs. Frank Brooks and Rucha Deshpande. 

\bibliography{ tmi}
\bibliographystyle{ieeetr}

\appendix

\subsection{Task-sensitive iFIM-map and task-insensitive iFIM-map for CNN}
\label{sec:AppendixA}

\begin{figure}[H]
  \centering
  \includegraphics[width=\linewidth]{ 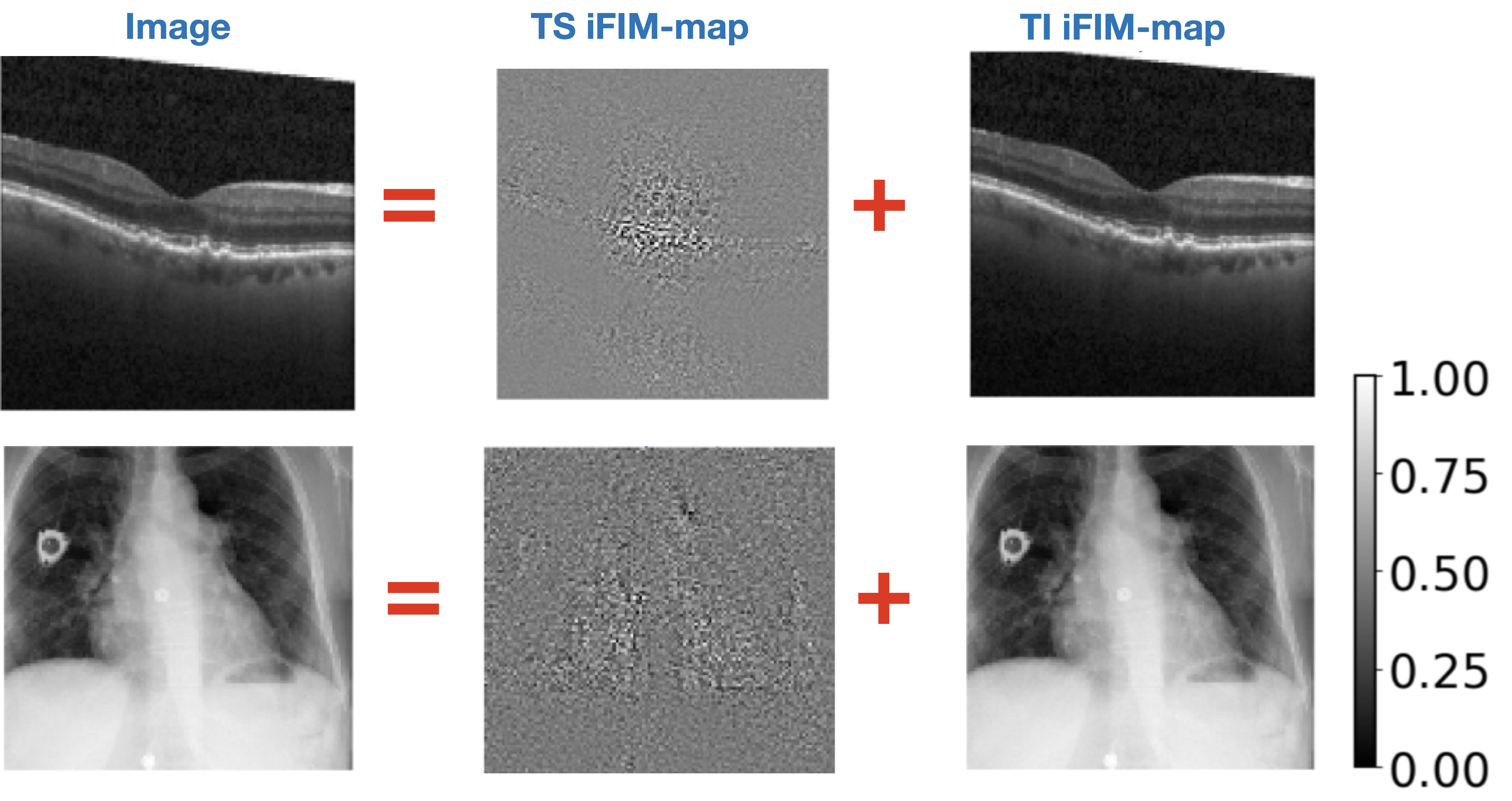} 
  \caption{Task-sensitive and task-insensitive iFIM-map for Drusen class of OCT classification task (top row) and Cardiomegaly class of chest X-ray classification task (bottom row).}
  \label{fig:appendix_a_fig}
\end{figure}

Similar to Sec.\ IV.C, additional task-sensitive and task-insensitive iFIM-maps for the OCT and chest X-ray classification tasks for shallow CNN classifier are shown in Fig.\ \ref{fig:appendix_a_fig}. Similar to the SKS/BKS example, the task-sensitive component highlights pixels that are more relevant to the model prediction, while the task-insensitive component captures comparatively less relevant pixels.

\subsection{Qualitative comparison with CAM-based post-hoc methods}

\begin{figure}[H]
    \centering
    \includegraphics[width=\linewidth]{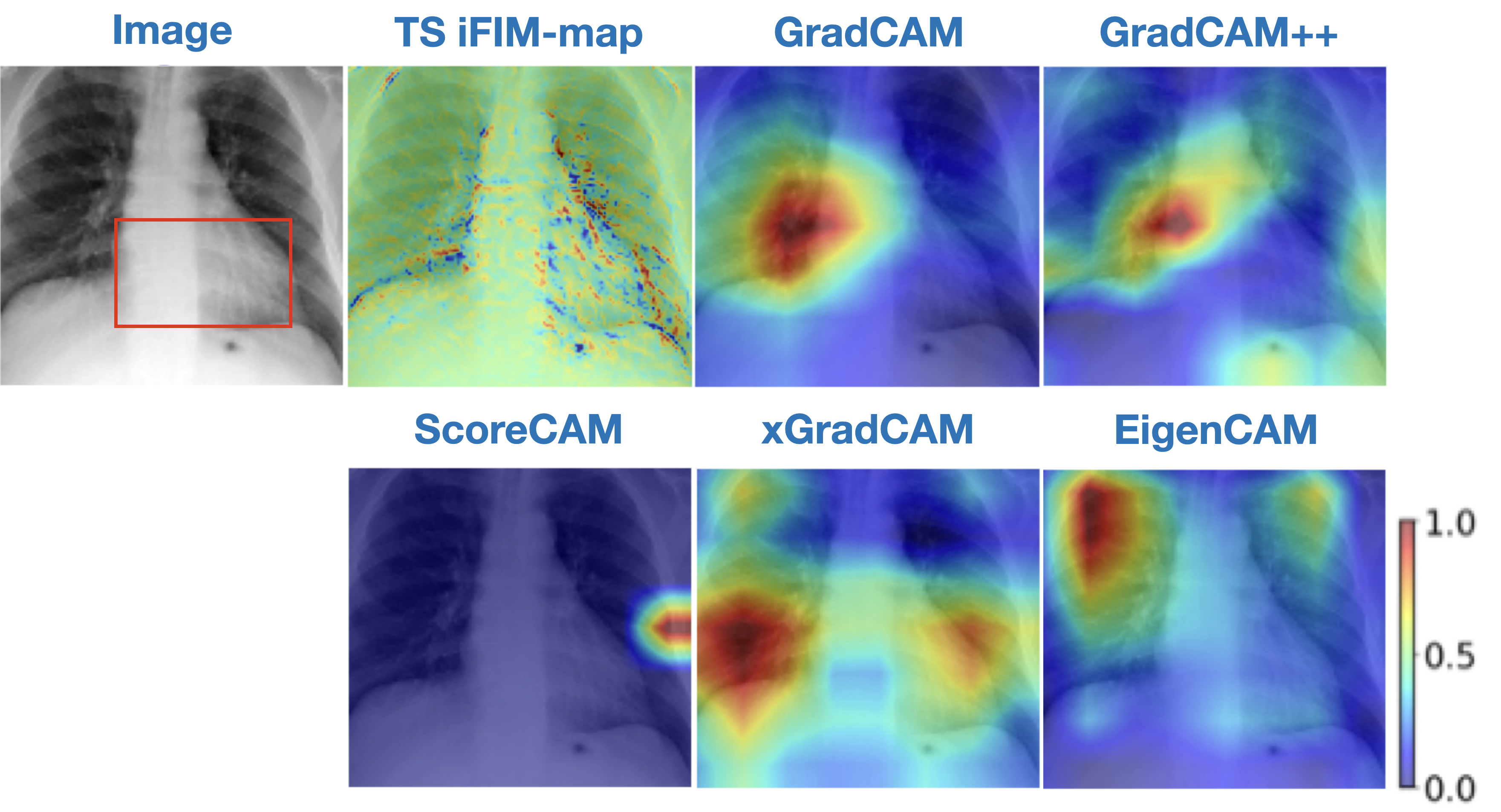}
    \caption{Qualitative comparison of five CAM-based attribution methods---Grad-CAM, Grad-CAM++, Score-CAM, XGrad-CAM, and Eigen-CAM---on a representative cardiomegaly case. Although all methods are applied to the same trained model and input image, the resulting attribution maps exhibit substantial variability in both localization and spatial extent. While some methods partially highlight regions near the cardiac silhouette, others produce diffuse or misplaced activations, including attention on irrelevant anatomical regions such as lung fields or image boundaries. The maps are overlaid on the original images using a contrast-adjusted jet colormap for visualization.}
    \label{fig:cam_cardiomegaly}
\end{figure}

Although the proposed iFIM-based framework is not directly comparable to conventional post-hoc attribution methods, a qualitative example is presented to illustrate how commonly used CAM-based techniques can yield substantially different attribution patterns for the same input and trained model. Specifically, a representative case from a cardiomegaly classification task is considered, where the presence of an enlarged cardiac silhouette is the defining diagnostic feature. Five widely used CAM-based methods---Grad-CAM, Grad-CAM++, Score-CAM, XGrad-CAM, and Eigen-CAM---are applied to the same trained model and input image. The resulting attribution maps are shown in Fig.~\ref{fig:cam_cardiomegaly}.

As observed in Fig.~\ref{fig:cam_cardiomegaly}, the different CAM-based methods produce substantially varying attribution patterns for the same input. While some methods partially highlight regions near the cardiac silhouette, others exhibit diffuse or misplaced activations, including attention to irrelevant anatomical structures such as lung fields or image boundaries. This example illustrates the variability that can arise across attribution methods even under identical model and data conditions. Although not directly comparable to the proposed framework, such variability motivates the use of complementary analysis tools that characterize model behavior through different quantities.
\end{document}